\documentclass[cameraready]{Interspeech}

\usepackage{booktabs, multirow} 
\usepackage{graphicx}
\usepackage{amsmath, amssymb}
\usepackage{url}

\usepackage{algorithm}
\usepackage{algpseudocode}
\title{RobustSpeechFlow: Learning Robust Text-to-Speech Trajectories via Augmentation-based Contrastive Flow Matching}

\author[affiliation={1}]{Jinhyeok}{Yang}
\author[affiliation={2}]{Hyeongju}{Kim$^\dagger$}
\author[affiliation={1}]{Yechan}{Yu}
\author[affiliation={1}]{Joon}{Byun}
\author[affiliation={1}]{Frederik}{Bous}
\author[affiliation={1}]{Juheon}{Lee}

\address{
    $^1$ Supertone Inc., Republic of Korea \\
    $^2$ Out of Set Inc., Republic of Korea
}
\email{hello@yangyangii.world, juheon@supertone.ai}

\keywords{text-to-speech, zero-shot TTS, flow matching, alignment robustness, contrastive learning}

\begin{document}
\maketitle
\let\thefootnote\relax\footnotetext{$^\dagger$This work was conducted while the author was with Supertone Inc.}

\begin{abstract}
While flow-matching text-to-speech (TTS) achieves strong zero-shot speaker similarity and naturalness, it remains susceptible to content fidelity issues, particularly skip and repeat errors from imperfect alignment. We propose RobustSpeechFlow, a training strategy that improves alignment robustness by extending contrastive flow matching with length-preserving repeat and skip latent augmentations. Requiring no external aligners or preference data, our method directly penalizes realistic failure modes and readily integrates into existing pipelines. On Seed-TTS-eval, it reduces the word error rate (WER) from 1.44 to 1.38 using only 0.06B parameters. On our ZERO500 benchmark, it delivers consistent intelligibility improvements across diverse speaker and prosody conditions; at NFE=24, it reduces English character error rate (CER) from 0.48\% to 0.35\% and Korean CER from 0.81\% to 0.57\%. Audio samples: \url{https://robustspeechflow.github.io/}.
\end{abstract}

\section{Introduction}
In text-to-speech (TTS), the primary requirement is \emph{content fidelity}: the system must render the intended text accurately. In production, failures such as word or phrase \emph{repetition} and \emph{skipping} are not minor artifacts; they materially reduce reliability and can create safety and compliance risks. As modern generative TTS systems continue to improve naturalness and expressivity, robust text--speech alignment remains a central problem.

Recent progress in non-autoregressive TTS has increasingly been driven by flow-matching-based latent generation. DiTTo-TTS shows that large-scale diffusion transformers can achieve strong zero-shot quality without relying on domain-specific factors such as phonemes or phoneme durations, highlighting the value of scalable latent modeling and cross-attention-based alignment \cite{ditto2025}. F5-TTS further demonstrates that flow matching with a DiT backbone can simplify the pipeline by removing explicit duration modeling and alignment modules while retaining high quality and robustness \cite{f5tts2025}. At the efficiency end of the spectrum, SupertonicTTS shows that compact flow-matching TTS can be practical for deployment, but also makes clear that content stability becomes increasingly difficult under tight capacity and latency constraints \cite{supertonic2025}.

In parallel, the integration of autoregressive language modeling with diffusion processes \cite{sun2024multimodal} has catalyzed a new wave of highly expressive speech generation. Building upon this paradigm, recent systems such as DiTAR \cite{ditar2025}, VoxCPM \cite{voxcpm2025}, and VibeVoice \cite{vibevoice2026} have substantially improved zero-shot quality and contextual naturalness. While these advances broaden the scope of generative speech modeling, they do not reliably resolve the issue of content fidelity. In practice, both pure flow-matching architectures and hybrid LM-diffusion systems continue to exhibit alignment errors. Crucially, these skip and repeat failures tend to become more severe as model capacity is constrained or the number of function evaluations (NFE) is reduced during inference \cite{f5tts2025,supertonic2025,ditar2025}. Furthermore, as speech generative modeling is scaled with increasingly large and diverse ASR-curated corpora such as GigaSpeech and Emilia \cite{gigaspeech2021,emilia2024}, ensuring robust quality control and alignment becomes even more critical.

Several directions have been explored to mitigate this misalignment. At the architectural level, explicit conditioning and guidance can improve low-step stability; for instance, DualSpeech uses a phoneme-level latent diffusion formulation with multiple classifier-free guidance signals to stabilize generation at low NFE \cite{dualspeech2024}. Similarly, stronger architectures and pretrained representations can improve overall robustness, but they do not eliminate skip and repeat failures altogether \cite{ditto2025}. Beyond architectural modifications, preference-based approaches construct dedicated intelligibility datasets and apply post-training objectives such as DPO to penalize failure cases \cite{intp2025,rafailov2023dpo}. Other approaches introduce auxiliary supervision from recognition or metric-optimization models, for example ASR- or CTC-derived objectives that directly penalize content errors \cite{dmdspeech2024}. While effective, these strategies either require costly data curation or introduce additional models and training complexity, which is undesirable for lightweight deployment settings.

We propose \textbf{RobustSpeechFlow}, a simple \emph{training strategy} for \emph{contrastive flow matching in TTS} that improves alignment robustness without additional models or datasets. Our key idea is to convert common TTS failure modes into \emph{hard negatives} in latent space: we synthesize skip- and repeat-like target latents with length-preserving overwrites and train the model contrastively so that the correct latent trajectory is preferred over these corrupted alternatives. This directly aligns the contrastive signal with the failure patterns that matter in practice while remaining compatible with standard flow-matching training.

The main contributions are as follows.
\begin{itemize}
    \item We introduce RobustSpeechFlow, a speech-specific contrastive flow-matching training strategy that targets skip/repeat failures via simple latent-space augmentations.
    \item The proposed method requires no preference dataset, no external aligner or ASR model, and readily integrates into FM-TTS pipelines with minimal overhead.
    \item We demonstrate consistent improvements on both a public zero-shot benchmark and a new multilingual benchmark with higher speaker, prosody, and text diversity, especially under low-NFE inference.
\end{itemize}

\section{Background}
\subsection{Alignment Robustness}
Alignment errors have been a persistent challenge since early deep-learning-based TTS. Tacotron~2 improves attention stability with location-sensitive attention \cite{shen2018tacotron2}. Deep Convolutional TTS (DCTTS) introduces a guided-attention loss and a forced-incremental attention heuristic at synthesis time to mitigate skipped or repeated characters \cite{tachibana2017guidedattention}. Despite these efforts, skip/repeat failures remain common, especially for long or out-of-distribution text.

Recent work has explored post-training and auxiliary-guidance approaches. Preference-based methods construct targeted failure cases and apply direct preference optimization (DPO) training \cite{intp2025,rafailov2023dpo}. Other approaches introduce auxiliary objectives, such as CTC-based ASR supervision, to directly penalize content errors \cite{dmdspeech2024,graves2006ctc}. Alternatively, leveraging strong pretrained text and speech representations (e.g., ByT5 and SpeechT5) has been shown to inherently improve alignment stability, as demonstrated in DiTTo-TTS \cite{ditto2025}. 

While positional encodings like LARoPE \cite{larope2025} provide a practical lever to improve alignment under duration variation, RobustSpeechFlow operates orthogonally: we reshape training using failure-mode contrastive negatives without introducing new models or external supervision.

\subsection{Contrastive Learning and Hard Negatives}
Contrastive learning has been widely used to improve discriminative representations by bringing positives closer and pushing negatives apart \cite{chen2020simclr,he2020moco}. In practice, the usefulness of the contrastive signal depends strongly on negative construction: hard negatives that remain close to the anchor but violate the conditioning semantics often provide a stronger gradient than easy, unrelated negatives.

This idea has recently been extended to generative modeling. Contrastive Flow Matching (CFM) \cite{cfm2025}, originally introduced for image generation, augments the standard conditional flow-matching objective with contrastive regularization terms computed from mismatched conditions. Concretely, it keeps the predicted vector field aligned with the positive transport direction while separating it from negative transport directions induced by incorrect conditions. This improves condition selectivity during training without changing the FM inference procedure. 

RobustSpeechFlow adapts this framework to speech generation and fundamentally strengthens the negative formulation for TTS. Rather than relying on random condition mismatches, we construct same-utterance latent corruptions that remain acoustically plausible while explicitly breaking local text--speech alignment.

\section{RobustSpeechFlow}
\subsection{Preliminaries: Conditional Flow Matching}
Let $x\in\mathbb{R}^{C\times T}$ be a continuous latent speech sequence produced by the \emph{Supertonic speech autoencoder} \cite{supertonic2025}, and let $c$ denote conditioning inputs (text and optional speaker prompt). We use a standard linear probability path for conditional flow matching:
\begin{equation}
\epsilon\sim\mathcal{N}(0,I),\quad t\sim\mathrm{U}(0,1),\quad x_t=(1-t)\epsilon+t x.
\end{equation}
The target velocity is $v(x,\epsilon)=x-\epsilon$. We train a neural vector field $u_\theta(x_t,t,c)$ with
\begin{equation}
\mathcal{L}_{\text{pos}}=\mathbb{E}\big[\lVert u_\theta(x_t,t,c)-v(x,\epsilon)\rVert_2^2\big].
\end{equation}

\subsection{Baseline: Contrastive FM with Random Negatives}
A simple contrastive extension samples a negative latent $x^-_{\text{rand}}$ from another batch item and introduces a contrastive regularization term based on the corresponding negative velocity:
\begin{equation}
\mathcal{L}_{\text{rand}}=\mathbb{E}\big[\lVert u_\theta(x_t,t,c)-v(x^-_{\text{rand}},\epsilon)\rVert_2^2\big].
\end{equation}
This term is subtracted in the final objective. In TTS, random negatives may be semantically unrelated to the conditioning text and thus fail to target alignment errors directly.

\subsection{TTS Failure-mode Negatives via Latent Augmentation}
TTS systems that lose alignment often repeat a pronunciation, word, or short clause, or skip a local text span altogether. We therefore construct \emph{TTS failure-mode negatives} by perturbing the \emph{ground-truth} latent sequence while keeping the overall length fixed. This design is important in practice: variable-length corruption complicates batch construction and destabilizes training in a fixed-length latent pipeline.

For each utterance, we first choose the augmentation type $m\in\{\text{rep},\text{skip}\}$ with $\Pr(m=\text{rep})=\Pr(m=\text{skip})=0.5$. We then sample a coverage budget $\kappa$: $\kappa\sim \mathrm{U}(0.2,0.4)$ for repeat and $\kappa\sim \mathrm{U}(0.4,0.8)$ for skip. Within the chosen mode, we repeatedly sample span lengths uniformly between the frame counts corresponding to 0.1 and 5.0 seconds and apply edits until the cumulative modified coverage would exceed the sampled budget.

\subsubsection{Repeat Augmentation} We first initialize the negative latent as a copy of the original sequence, $x^-_{\text{rep}} = x$. We then sample a source span $[s,s+\ell)$ and a different target start $k$ with $k+\ell\leq T$, and overwrite the target region with the source content:
\begin{equation}
x^-_{\text{rep}}[k:k+\ell] \leftarrow x[s:s+\ell], \quad \text{s.t. } s \neq k.
\end{equation}
By overwriting a target region rather than appending duplicated frames, this approach strictly preserves the original sequence length. Moreover, because the original content at the target span is overwritten, this operation inherently induces a simultaneous skip error, effectively simulating the composite repeat-and-skip failures frequently observed in real-world TTS misalignment.

\subsubsection{Skip Augmentation} Similarly, we initialize the negative latent as a copy of the original sequence, $x^-_{\text{skip}} = x$. We sample a skip start index $s_1$ and a skip length $\ell$. To simulate a local skip, we shift the subsequent latent sequence forward to overwrite the skipped region:
\begin{equation}
x^-_{\text{skip}}[s_1:T-\ell] \leftarrow x[s_1+\ell:T].
\end{equation}
After shifting the sequence, the remaining $\ell$ frames at the tail of the utterance ($[T-\ell:T]$) are replaced with a precomputed silence latent $x_{\text{sil}}$. By shifting the future context and padding with silence rather than explicitly deleting frames, this approach simulates a realistic skip error while strictly preserving the global sequence duration.

\subsubsection{Implementation Details} These negatives remain close to the original utterance in speaker identity and acoustic texture, but they corrupt local text--speech correspondence. They are therefore substantially harder and more relevant to the conditioning text than random batch negatives. For implementation convenience, the silence latent $x_{\text{sil}}$ used to pad the tail of the skip negatives is simply precomputed by encoding a zero-padded waveform with the codec and repeating the resulting latent frames to the required length.

The augmentation-based contrastive regularization is defined as:
\begin{equation}
\mathcal{L}_{\text{aug}}=\mathbb{E}\big[\lVert u_\theta(x_t,t,c)-v(x^-_{\text{aug}},\epsilon)\rVert_2^2\big],
\end{equation}
where $x^-_{\text{aug}} \in \{x^-_{\text{rep}}, x^-_{\text{skip}}\}$ denotes the corrupted latent produced by the chosen augmentation mode.

\subsection{Overall Objective}
The overall training objective combines the positive flow-matching loss with the contrastive regularization terms:
\begin{equation}
\mathcal{L}=\mathcal{L}_{\text{pos}}-\lambda_{\text{rand}}\,\mathcal{L}_{\text{rand}}-\lambda_{\text{aug}}\,\mathcal{L}_{\text{aug}}.
\end{equation}

\begin{algorithm}[t]
\caption{Training step of RobustSpeechFlow}
\label{alg:robust_flow}
\begin{algorithmic}[1]
\Require Batch of audio $a$, Text conditions $c$, Hyperparameters $\lambda_{\text{rand}}$, $\lambda_{\text{aug}}$
\State $x \gets \text{Encoder}(a)$ \Comment{Extract speech latent representation}
\State $\epsilon \sim \mathcal{N}(0, I)$,\quad $t \sim \mathcal{U}(0, 1)$ \Comment{Sample Gaussian noise and time step}
\State $x_t \gets (1-t)\epsilon + t x$ \Comment{Construct flow matching trajectory}
\State $v_{\text{pos}} \gets x - \epsilon$ \Comment{Target vector field (Positive)}
\State $x_{\text{rand}} \gets \text{RollBatch}(x)$ \Comment{In-batch random negative sample}
\State $v_{\text{rand}} \gets x_{\text{rand}} - \epsilon$ \Comment{Vector field towards random negative}
\State $x_{\text{aug}} \gets \text{Augment}(x)$ \Comment{Same-length failure-mode negative}
\State $v_{\text{aug}} \gets x_{\text{aug}} - \epsilon$ \Comment{Vector field towards augmented negative}
\State $u_\theta \gets \text{Model}(x_t, t, c)$ \Comment{Predict vector field}
\State $\mathcal{L}_{\text{pos}} \gets \lVert u_\theta - v_{\text{pos}} \rVert_2^2$ \Comment{Standard FM loss}
\State $\mathcal{L}_{\text{rand}} \gets \lVert u_\theta - v_{\text{rand}} \rVert_2^2$ \Comment{Contrastive reg. (Random)}
\State $\mathcal{L}_{\text{aug}} \gets \lVert u_\theta - v_{\text{aug}} \rVert_2^2$ \Comment{Contrastive reg. (Hard)}
\State $\mathcal{L}_{\text{total}} \gets \mathcal{L}_{\text{pos}} - \lambda_{\text{rand}} \mathcal{L}_{\text{rand}} - \lambda_{\text{aug}} \mathcal{L}_{\text{aug}}$ \Comment{Total objective}
\State \textbf{return} $\mathcal{L}_{\text{total}}$ \Comment{Update model parameters via gradient descent}
\end{algorithmic}
\end{algorithm}

\section{Experimental Setup}

\subsection{Training Data}
We train on internal corpora of approximately 10k hours, 5M utterances, and 80k speakers per language (English and Korean), utilizing a mix of human-annotated and ASR-generated transcriptions.

\subsection{Model and Baselines}
We apply RobustSpeechFlow to SupertonicTTS \cite{supertonic2025}, a compact flow-matching model operating on \emph{Supertonic autoencoder} latents. Keeping the architecture fixed, we compare three objectives: (i) \emph{Baseline} (vanilla SupertonicTTS), (ii) \emph{ContrastiveFM} (baseline with random-batch negatives), and (iii) \emph{RobustSpeechFlow} (ContrastiveFM with augmentation-based hard negatives). To isolate the objective's effect, we train an utterance-level duration predictor independently and share the exact same pretrained text-to-latent checkpoint across all methods. Following SupertonicTTS, we input raw text without grapheme-to-phoneme conversion.

\subsection{Training and Inference Settings}
Audio is resampled to 44.1 kHz for both training and inference. We train all models for 500k steps on 8 NVIDIA H100 GPUs using dynamic batching. We use the AdamW optimizer (learning rate 5e-4, $\beta=(0.9, 0.999)$, zero weight decay), halving the learning rate every 200k steps. Reference speech is sampled from a same-speaker segment uniformly drawn from 3 to 10 seconds. For all methods, we apply Length-Aware RoPE \cite{larope2025} and context-sharing batch expansion (factor 6) \cite{supertonic2025}. While RobustSpeechFlow uses $\lambda_{\text{rand}}=\lambda_{\text{aug}}=0.2$, ContrastiveFM sets $\lambda_{\text{aug}}=0.0$. For deployment-oriented inference, we use an Euler solver with a classifier-free guidance weight of 3.0 at NFE $\in \{12,24\}$.

\subsection{Evaluation Metrics}
We evaluate on two suites: the public Seed-TTS-eval benchmark \cite{seedtts2024} and our newly constructed ZERO500 benchmark. ZERO500 is designed to stress alignment under diverse conditions, containing 50 diverse reference voices (e.g., game, news, conversational) per language (English and Korean). Each voice is randomly paired with 10 text prompts, yielding 500 pairs per language. We synthesize each pair twice using different random seeds and report the average.

We transcribe the synthesized audio using Whisper large-v3 \cite{radford2022whisper} to compute the character error rate (CER, \%) and word error rate (WER, \%). Scoring applies only light text normalization, limited to punctuation removal and simple whitespace cleanup.

\section{Results}

\begin{table}[t]
  \caption{Performance on the Seed-TTS-eval benchmark. Lower WER and higher SIM are better. The lower block reports our compact in-family baselines and RobustSpeechFlow.}
  \label{tab:seed_tts_eval_en}
  \centering
  \footnotesize
  \setlength{\tabcolsep}{3pt}
  \begin{tabular}{@{}p{0.52\columnwidth}ccc@{}}
    \toprule
    \textbf{Model} & \textbf{Params} & \textbf{WER}$\downarrow$ & \textbf{SIM}$\uparrow$ \\
    \midrule
    MegaTTS3 \cite{megatts3_2025} & 0.5B & 2.79 & \underline{0.77} \\
    Seed-TTS$_{DiT}$ \cite{seedtts2024} & -- & 1.73 & \textbf{0.79} \\
    DiTAR \cite{ditar2025} & 0.6B & 1.69 & 0.74 \\
    MiniMax-Speech \cite{minimaxspeech2025} & -- & 1.65 & 0.69 \\
    F5-TTS \cite{f5tts2025} & 0.3B & 2.00 & 0.67 \\
    CosyVoice3 \cite{cosyvoice3_2025} & 1.5B & 2.22 & 0.72 \\
    Spark-TTS \cite{sparktts2025} & 0.5B & 3.14 & 0.57 \\
    OpenAudio S1-Mini \cite{openaudios1_2025} & 0.5B & 1.94 & 0.55 \\
    IndexTTS2 \cite{indextts2_2025} & 1.5B & 2.23 & 0.71 \\
    VibeVoice \cite{vibevoice2026} & 1.5B & 3.04 & 0.69 \\
    VoxCPM-Emilia \cite{voxcpm2025} & 0.5B & 2.34 & 0.68 \\
    VoxCPM \cite{voxcpm2025} & 0.5B & 1.85 & 0.73 \\
    \midrule
    Baseline(SupertonicTTS) & \textbf{0.06B} & 1.44 & 0.60 \\
    ContrastiveFM & \textbf{0.06B} & \underline{1.41} & 0.60 \\
    \textbf{RobustSpeechFlow} & \textbf{0.06B} & \textbf{1.38} & 0.60 \\
    \bottomrule
  \end{tabular}
\end{table}

\subsection{Zero-shot Intelligibility on Seed-TTS-eval}
We first evaluate on the public Seed-TTS-eval benchmark \cite{seedtts2024}. Table~\ref{tab:seed_tts_eval_en} summarizes the benchmark comparison against representative zero-shot TTS systems together with our in-family baselines. Within the compact SupertonicTTS setup, standard ContrastiveFM already reduces the WER from 1.44 to 1.41, and RobustSpeechFlow further improves it to 1.38 while keeping the same similarity score (SIM) of 0.60. This corresponds to a relative WER reduction of 4.2\% over the vanilla baseline and 2.1\% over standard ContrastiveFM. Remarkably, this achieves the lowest WER across the entire benchmark despite utilizing only 0.06B parameters. These results show that alignment-focused contrastive training can materially improve intelligibility even in a compact model regime. Notably, many competing systems are 5 to over 20 times larger, yet do not match this WER. SIM remains unchanged across the three compact variants, suggesting that the WER gain is strictly attributable to improved alignment rather than a shift in speaker conditioning.

\begin{table}[t]
  \caption{Results on ZERO500 at 500k steps (\%). Lower is better.}
  \label{tab:zero500_main}
  \centering
  \footnotesize
  \setlength{\tabcolsep}{4pt} %
  \begin{tabular}{lcc cc cc}
    \toprule
    \multirow{2}{*}{\textbf{Model}} & \multirow{2}{*}{\textbf{NFE}} & \multicolumn{2}{c}{\textbf{English}} & \multicolumn{2}{c}{\textbf{Korean}}\\
    \cmidrule(lr){3-4} \cmidrule(lr){5-6}
     &  & \textbf{CER}$\downarrow$ & \textbf{WER}$\downarrow$ & \textbf{CER}$\downarrow$ & \textbf{WER}$\downarrow$ \\
    \midrule
    Baseline & 12 & 0.55 & 1.25 & 0.93 & 8.46 \\
    Baseline & 24 & 0.48 & 1.18 & 0.81 & 8.40 \\
    ContrastiveFM & 12 & 0.41 & 1.10 & 0.77 & 7.92 \\
    ContrastiveFM & 24 & \underline{0.39} & \underline{1.06} & \underline{0.65} & 7.72 \\
    \textbf{RobustSpeechFlow} & 12 & 0.43 & 1.14 & \textbf{0.57} & \underline{7.59} \\
    \textbf{RobustSpeechFlow} & 24 & \textbf{0.35} & \textbf{1.03} & \textbf{0.57} & \textbf{7.45} \\
    \bottomrule
  \end{tabular}
\end{table}

\subsection{Alignment Robustness in Diverse Conditions}
While Seed-TTS-eval is a standard zero-shot benchmark, its reliance on read speech with limited prosodic variation restricts its ability to reflect diverse real-world usage conditions. To better evaluate alignment robustness under such diversity, we constructed the ZERO500 benchmark for English and Korean and evaluated all systems at the 500k training step.

Table~\ref{tab:zero500_main} demonstrates that RobustSpeechFlow consistently outperforms the baselines, providing the most reliable alignment. Specifically, under the highly constrained NFE=12 setting, the proposed method reduces the Korean CER from 0.93\% to 0.57\% and the WER from 8.46\% to 7.59\% compared to the vanilla baseline. This strong performance holds at NFE=24, where CER is further reduced to 0.57\% and WER to 7.45\%. On the English subset, RobustSpeechFlow similarly improves upon the baseline across both NFE settings, ultimately achieving the best overall result at NFE=24 with a CER of 0.35\% and a WER of 1.03\%. Although standard ContrastiveFM shows slight advantages on English at NFE=12, its gains do not transfer as consistently across languages and NFE regimes. Overall, these metrics suggest that augmentation-based hard negatives provide a fundamentally more robust alignment signal than random utterance mismatches alone.

\subsection{Stability Across Training Stages}
\begin{figure}[t]
  \centering
  \begin{minipage}{0.22\textwidth}
    \centering
    \includegraphics[width=\linewidth]{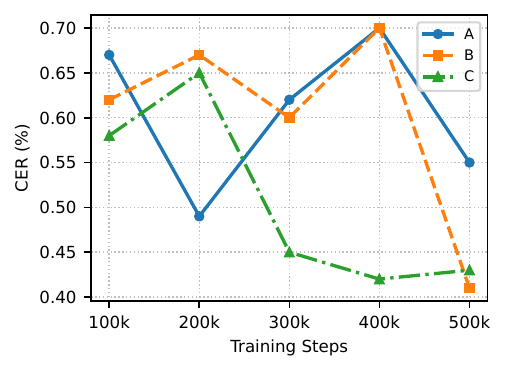}
    \\[-1mm]
    {\scriptsize (a) ZERO500-en CER (\%), NFE=12}
  \end{minipage}\hfill
  \begin{minipage}{0.22\textwidth}
    \centering
    \includegraphics[width=\linewidth]{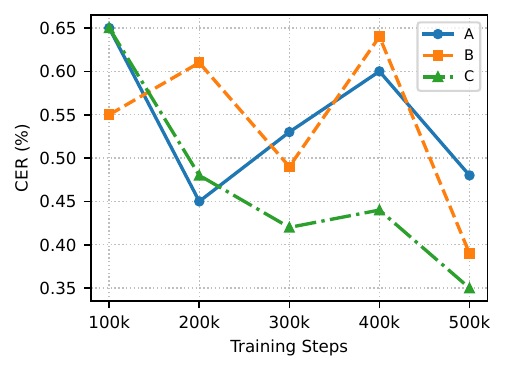}
    \\[-1mm]
    {\scriptsize (b) ZERO500-en CER (\%), NFE=24}
  \end{minipage}

  \vspace{1mm}

  \begin{minipage}{0.22\textwidth}
    \centering
    \includegraphics[width=\linewidth]{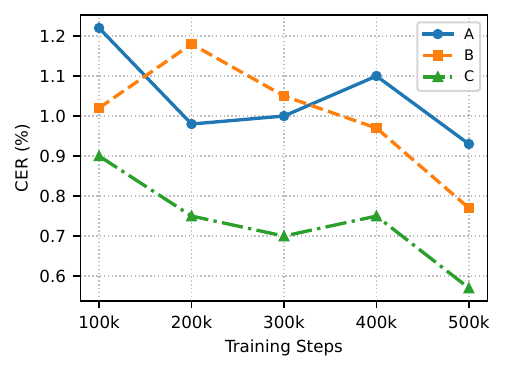}
    \\[-1mm]
    {\scriptsize (c) ZERO500-ko CER (\%), NFE=12}
  \end{minipage}\hfill
  \begin{minipage}{0.22\textwidth}
    \centering
    \includegraphics[width=\linewidth]{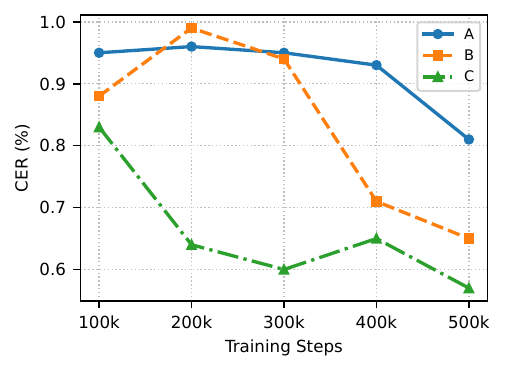}
    \\[-1mm]
    {\scriptsize (d) ZERO500-ko CER (\%), NFE=24}
  \end{minipage}
  \caption{CER (\%) over training steps on ZERO500. In each panel, legend A, B, and C correspond to Baseline, ContrastiveFM, and RobustSpeechFlow, respectively.}
  \label{fig:cer_curve}
\end{figure}

Figure~\ref{fig:cer_curve} provides a clearer perspective on how alignment stability evolves over the course of training, contextualizing the final results from Table~\ref{tab:zero500_main}. First, RobustSpeechFlow demonstrates markedly superior reliability on Korean. Across both NFE settings, its CER steadily drops and stabilizes at 0.57\% at 500k, whereas the baseline plateaus at higher error rates (0.81\% at NFE=24) and ContrastiveFM ends at 0.65\%. This highlights the critical benefit of failure-mode negatives when the target language and benchmark contain high prosodic variation.

On English, the training dynamics present a more complex picture. While standard ContrastiveFM is competitive early on, its trajectory exhibits higher variance. Conversely, RobustSpeechFlow establishes a more consistent optimization path as training progresses, notably overtaking all methods from 300k onward at NFE=24 to reach the lowest final CER. We hypothesize that explicitly penalizing the model for entering latent regions associated with skip and repeat errors effectively stabilizes the loss landscape for cross-attention alignment during the later stages of training. Consequently, augmentation-based negatives yield the most consistent gains in high-stress alignment scenarios, particularly for low-NFE regimes.

\section{Discussion and Conclusion}
We presented RobustSpeechFlow, a zero-shot TTS training strategy that improves content fidelity through length-preserving repeat and skip augmentations in contrastive flow matching. By simulating these common failure modes within the latent space, our approach directly aligns the contrastive penalty with the structured errors observed in real-world deployment. Consequently, without requiring external models or preference data, it reduces the WER on Seed-TTS-eval from 1.44 (baseline) to 1.38 using only 0.06B parameters. While standard contrastive flow matching can remain competitive in certain specific settings, our augmentation strategy demonstrates broader reliability and delivers the most consistent intelligibility improvements on the diverse ZERO500 benchmark, particularly under low-NFE regimes.

While RobustSpeechFlow excels in alignment robustness, a trade-off with speaker similarity remains on the public benchmark. We hypothesize that this gap stems primarily from the representational limits of the compact baseline architecture rather than the proposed objective itself, and can be mitigated by scaling the speech codec and backbone. Furthermore, while objective ASR-based metrics are valuable for scalable benchmarking, they can be inherently biased by recognition errors and arbitrary text normalization choices. Future work will complement these metrics with comprehensive subjective assessments (e.g., MOS), expand the design of speech-mimicking negative latents to target a broader taxonomy of realistic failure modes, and investigate the generalizability of this approach across larger flow-matching and autoregressive-diffusion frameworks.

\clearpage
\section{Generative AI Use Disclosure}
Generative AI tools were used exclusively for proofreading, improving the grammatical quality of the manuscript, and assisting in the preparation of the demo webpage. The core text and all scientific contributions were fully developed and authored by the human authors.

\bibliographystyle{IEEEtran}
\bibliography{reference}

@inproceedings{shen2018tacotron2,
  title     = {Natural TTS Synthesis by Conditioning WaveNet on Mel Spectrogram Predictions},
  author    = {Shen, Jonathan and Pang, Ruoming and Weiss, Ron J. and Schuster, Mike and Jaitly, Navdeep and Yang, Zongheng and Chen, Zhifeng and Zhang, Yu and Wang, Yuxuan and Skerry-Ryan, R. J. and Saurous, Rif A. and Agiomyrgiannakis, Yannis and Wu, Yonghui},
  booktitle = {Proc. IEEE ICASSP},
  year      = {2018}
}

@inproceedings{tachibana2017guidedattention,
  title     = {Efficiently Trainable Text-to-Speech System Based on Deep Convolutional Networks with Guided Attention},
  author    = {Tachibana, Hideyuki and Uenoyama, Katsuya and Aihara, Shunsuke},
  booktitle = {Proc. IEEE ICASSP},
  pages     = {4784--4788},
  year      = {2018}
}

@inproceedings{graves2006ctc,
  title     = {Connectionist Temporal Classification: Labelling Unsegmented Sequence Data with Recurrent Neural Networks},
  author    = {Graves, Alex and Fern{\'a}ndez, Santiago and Gomez, Faustino and Schmidhuber, J{\"u}rgen},
  booktitle = {Proceedings of the 23rd International Conference on Machine Learning},
  pages     = {369--376},
  year      = {2006}
}

@inproceedings{chen2020simclr,
  title     = {A Simple Framework for Contrastive Learning of Visual Representations},
  author    = {Chen, Ting and Kornblith, Simon and Norouzi, Mohammad and Hinton, Geoffrey},
  booktitle = {Proceedings of the 37th International Conference on Machine Learning},
  pages     = {1597--1607},
  year      = {2020}
}

@inproceedings{he2020moco,
  title     = {Momentum Contrast for Unsupervised Visual Representation Learning},
  author    = {He, Kaiming and Fan, Haoqi and Wu, Yuxin and Xie, Saining and Girshick, Ross},
  booktitle = {Proc. IEEE/CVF Conference on Computer Vision and Pattern Recognition},
  pages     = {9729--9738},
  year      = {2020}
}

@inproceedings{cfm2025,
  title     = {Contrastive Flow Matching},
  author    = {Stoica, George and Ramanujan, Vivek and Fan, Xiang and Farhadi, Ali and Krishna, Ranjay and Hoffman, Judy},
  booktitle = {Proceedings of the IEEE/CVF International Conference on Computer Vision (ICCV)},
  year      = {2025}
}

@article{supertonic2025,
  title={SupertonicTTS: Towards Highly Efficient and Streamlined Text-to-Speech System},
  author={Kim, Hyeongju and Yang, Jinhyeok and Yu, Yechan and Ji, Seunghun and Morton, Jacob and Bous, Frederik and Byun, Joon and Lee, Juheon},
  journal={arXiv preprint arXiv:2503.23108},
  year={2025}
}

@article{larope2025,
  title={Length-aware rotary position embedding for text-speech alignment},
  author={Kim, Hyeongju and Lee, Juheon and Yang, Jinhyeok and Morton, Jacob},
  journal={arXiv preprint arXiv:2509.11084},
  year={2025}
}

@inproceedings{ditto2025,
  title={DiTTo-TTS: Diffusion Transformers for Scalable Text-to-Speech without Domain-Specific Factors},
  author={Lee, Keon and Kim, Dong Won and Kim, Jaehyeon and Chung, Seungjun and Cho, Jaewoong},
  booktitle = {International Conference on Learning Representations (ICLR)},
  year      = {2025}
}

@inproceedings{dmdspeech2024,
  title     = {DMDSpeech: Distilled Diffusion Model Surpassing the Teacher in Zero-shot Speech Synthesis via Direct Metric Optimization},
  author    = {Li, Yinghao Aaron and Kumar, Rithesh and Jin, Zeyu},
  booktitle = {International Conference on Learning Representations (ICLR)},
  year      = {2025}
}

@inproceedings{intp2025,
  title={Advancing zero-shot text-to-speech intelligibility across diverse domains via preference alignment},
  author={Zhang, Xueyao and Wang, Yuancheng and Wang, Chaoren and Li, Ziniu and Chen, Zhuo and Wu, Zhizheng},
  booktitle={Proceedings of the 63rd Annual Meeting of the Association for Computational Linguistics (Volume 1: Long Papers)},
  pages={12251--12270},
  year={2025}
}

@inproceedings{rafailov2023dpo,
  title     = {Direct Preference Optimization: Your Language Model is Secretly a Reward Model},
  author    = {Rafailov, Rafael and Sharma, Archit and Mitchell, Eric and Manning, Christopher D. and Ermon, Stefano and Finn, Chelsea},
  booktitle = {Advances in Neural Information Processing Systems},
  year      = {2023}
}

@inproceedings{dualspeech2024,
  title     = {DualSpeech: Enhancing Speaker-Fidelity and Text-Intelligibility Through Dual Classifier-Free Guidance},
  author    = {Yang, Jinhyeok and Lee, Junhyeok and Choi, Hyeong-Seok and Ji, Seunghun and Kim, Hyeongju and Lee, Juheon},
  booktitle = {Proc. Interspeech},
  year      = {2024}
}

@inproceedings{gigaspeech2021,
  title={GigaSpeech: An Evolving, Multi-Domain ASR Corpus with 10,000 Hours of Transcribed Audio},
  author={Chen, Guoguo and Chai, Shuzhou and Wang, Guan-Bo and Du, Jiayu and Zhang, Wei-Qiang and Weng, Chao and Su, Dan and Povey, Daniel and Trmal, Jan and Zhang, Junbo and others},
  booktitle={Proc. Interspeech 2021},
  pages={3670--3674},
  year={2021}
}

@article{emilia2024,
  title={Emilia: An extensive, multilingual, and diverse speech dataset for large-scale speech generation},
  author={He, Haorui and Shang, Zengqiang and Wang, Chaoren and Li, Xuyuan and Gu, Yicheng and Hua, Hua and Liu, Liwei and Yang, Chen and Li, Jiaqi and Shi, Peiyang and others},
  journal={2024 IEEE Spoken Language Technology Workshop (SLT)},
  pages={885--890},
  year={2024},
  organization={IEEE}
}

@article{seedtts2024,
  title   = {Seed-TTS: A Family of High-Quality Versatile Speech Generation Models},
  author  = {Anastassiou, Philip and Chen, Jiawei and Chen, Jitong and others},
  journal = {arXiv preprint arXiv:2406.02430},
  year    = {2024}
}

@article{megatts3_2025,
  title   = {MegaTTS 3: Sparse Alignment Enhanced Latent Diffusion Transformer for Zero-Shot Speech Synthesis},
  author  = {Jiang, Ziyue and Ren, Yi and Li, Ruiqi and Ji, Shengpeng and Zhang, Boyang and Ye, Zhenhui and Zhang, Chen and Bai, Jionghao and Yang, Xiaoda and Zuo, Jialong and Zhang, Yu and Liu, Rui and Yin, Xiang and Zhao, Zhou},
  journal = {arXiv preprint arXiv:2502.18924},
  year    = {2025}
}

@article{minimaxspeech2025,
  title   = {MiniMax-Speech: Intrinsic Zero-Shot Text-to-Speech with a Learnable Speaker Encoder},
  author  = {Zhang, Bowen and Guo, Congchao and Yang, Geng and Yu, Hang and Zhang, Haozhe and Lei, Heidi and Mai, Jialong and Yan, Junjie and Yang, Kaiyue and Yang, Mingqi and Huang, Peikai and Jin, Ruiyang and Jiang, Sitan and Cheng, Weihua and Li, Yawei and Xiao, Yichen and Zhou, Yiying and Zhang, Yongmao and Lu, Yuan and He, Yucen},
  journal = {arXiv preprint arXiv:2505.07916},
  year    = {2025}
}

@inproceedings{f5tts2025,
  title={F5-tts: A fairytaler that fakes fluent and faithful speech with flow matching},
  author={Chen, Yushen and Niu, Zhikang and Ma, Ziyang and Deng, Keqi and Wang, Chunhui and Zhao, Jian and Yu, Kai and Chen, Xie},
  booktitle={Proceedings of the 63rd Annual Meeting of the Association for Computational Linguistics (Volume 1: Long Papers)},
  pages={6255--6271},
  year={2025}
}

@article{cosyvoice3_2025,
  title   = {CosyVoice 3: Towards In-the-wild Speech Generation via Scaling-up and Post-training},
  author  = {Du, Zhihao and Gao, Changfeng and Wang, Yuxuan and Yu, Fan and Zhao, Tianyu and Wang, Hao and Lv, Xiang and Wang, Hui and Shi, Xian and An, Keyu and others},
  journal = {arXiv preprint arXiv:2505.17589},
  year    = {2025}
}

@article{sparktts2025,
  title   = {Spark-TTS: An Efficient LLM-Based Text-to-Speech Model with Single-Stream Decoupled Speech Tokens},
  author  = {Wang, Xinsheng and Jiang, Mingqi and Ma, Ziyang and Zhang, Ziyu and Liu, Songxiang and Li, Linqin and Liang, Zheng and Zheng, Qixi and Wang, Rui and Feng, Xiaoqin and others},
  journal = {arXiv preprint arXiv:2503.01710},
  year    = {2025}
}

@misc{openaudios1_2025,
  author       = {{OpenAudio}},
  title        = {Introducing S1},
  year         = {2025},
  howpublished = {OpenAudio technical blog},
  note         = {June 3, 2025}
}

@article{indextts2_2025,
  title   = {IndexTTS2: A Breakthrough in Emotionally Expressive and Duration-Controlled Auto-Regressive Zero-Shot Text-to-Speech},
  author  = {Zhou, Siyi and Zhou, Yiquan and He, Yi and Zhou, Xun and Wang, Jinchao and Deng, Wei and Shu, Jingchen},
  journal = {arXiv preprint arXiv:2506.21619},
  year    = {2025}
}

@inproceedings{vibevoice2026,
title={VibeVoice: Expressive Podcast Generation with Next-Token Diffusion},
author={Zhiliang Peng and Jianwei Yu and Wenhui Wang and Yaoyao Chang and Yutao Sun and Li Dong and Yi Zhu and Weijiang Xu and Hangbo Bao and Zehua Wang and Shaohan Huang and Yan Xia and Furu Wei},
booktitle={The Fourteenth International Conference on Learning Representations},
year={2026},
}

@article{voxcpm2025,
  title   = {VoxCPM: Tokenizer-Free TTS for Context-Aware Speech Generation and True-to-Life Voice Cloning},
  author  = {Zhou, Yixuan and Zeng, Guoyang and Liu, Xin and Li, Xiang and Yu, Renjie and Wang, Ziyang and Ye, Runchuan and Sun, Weiyue and Gui, Jiancheng and Li, Kehan and Wu, Zhiyong and Liu, Zhiyuan},
  journal = {arXiv preprint arXiv:2509.24650},
  year    = {2025}
}

@inproceedings{radford2022whisper,
  title     = {Robust Speech Recognition via Large-Scale Weak Supervision},
  author    = {Radford, Alec and Kim, Jong Wook and Xu, Tao and Brockman, Greg and McLeavey, Christine and Sutskever, Ilya},
  booktitle = {Proceedings of the 40th International Conference on Machine Learning},
  year      = {2023}
}

@article{ditar2025,
  title={Ditar: Diffusion transformer autoregressive modeling for speech generation},
  author={Jia, Dongya and Chen, Zhuo and Chen, Jiawei and Du, Chenpeng and Wu, Jian and Cong, Jian and Zhuang, Xiaobin and Li, Chumin and Wei, Zhen and Wang, Yuping and others},
  journal={arXiv preprint arXiv:2502.03930},
  year={2025}
}

@article{sun2024multimodal,
  title={Multimodal latent language modeling with next-token diffusion},
  author={Sun, Yutao and Bao, Hangbo and Wang, Wenhui and Peng, Zhiliang and Dong, Li and Huang, Shaohan and Wang, Jianyong and Wei, Furu},
  journal={arXiv preprint arXiv:2412.08635},
  year={2024}
}

\end{document}